\documentclass[11pt]{article}
\usepackage{moriond,epsfig,wrapfig,tabularx}

\def\Journal#1#2#3#4{{#1} {\bf #2}, #3 (#4)}

\def\APP{\em Astropart. Phys.}

\def\PRL{\em Phys. Rev. Lett.}
\def\PRD{{\em Phys. Rev.} D}

\def\be{\begin{equation}}
\def\ee{\end{equation}}
\def\bea{\begin{eqnarray}}
\def\eea{\end{eqnarray}}

\begin{document}
\vspace*{4cm}
\title{RECENT RESULTS FROM THE AMANDA EXPERIMENT}

\author{P. Niessen}

\address{IIHE, Vrije Universiteit Brussel, Pleinlaan 2,\\
 B-1050 Brussel, Belgium\\
 {\rm for the AMANDA COLLABORATION: }\\
%%
% add sloppypar command to prevent hyphenation
\begin{sloppypar}
\noindent
{\rm
J.~Ahrens$^{11}$, 
X.~Bai$^{1}$, 
%G.~Barouch$^{13}$, 
S.W.~Barwick$^{10}$, 
%R.C.~Bay$^{8}$, 
T.~Becka$^{11}$, 
J.K.~Becker$^{2}$,
K.-H.~Becker$^{2}$, 
E.~Bernardini$^{4}$,
D.~Bertrand$^{3}$, 
F.~Binon$^{3}$,
A.~Biron$^{4}$
% ...
%S.~Young$^{9}$
} %rm
\end{sloppypar}
\vspace*{0.2cm} 
{\footnotesize
\noindent
   (1) Bartol Research Institute, University of Delaware, Newark, DE 19716, USA
   \newline
   (2) Fachbereich 8 Physik, BUGH Wuppertal, D-42097 Wuppertal, Germany
   \newline
   (3) Universit\'e Libre de Bruxelles, Science Faculty, CP230, B-1050 Brussels, Belgium
   \newline
   (4) DESY-Zeuthen, D-15735 Zeuthen, Germany
   \newline
   (5) Blackett Laboratory, Imperial College, London SW7 2BW, UK
   \newline
   (6) Dept. of Technology, Kalmar University, S-39182 Kalmar, Sweden
   \newline
   (7) Lawrence Berkeley National Laboratory, Berkeley, CA 94720, USA
   \newline
   (8) Dept. of Physics, Pennsylvania State University, University Park, PA 16802, USA
   \newline
   (9) Dept. of Physics, University of California, Berkeley, CA 94720, USA
   \newline
   (10) Dept. of Physics and Astronomy, University of California, Irvine, CA 92697, USA
   \newline
   (11) Institute of Physics, University of Mainz, D-55099 Mainz, Germany
   \newline
   (12) Dept. of Physics, University of Maryland, College Park, MD 20742, USA
   \newline
   (13) University of Mons-Hainaut, 7000 Mons, Belgium
   \newline
   (14) Departamento de F\'{\i}sica, Universidad Sim\'on Bol\'{\i}var, Caracas,  1080 Venezuela
   \newline
   (15) Dept. of Physics, University of Wisconsin, Madison, WI 53706, USA
   \newline
   (16) Physics Dept., University of Wisconsin, River Falls, WI 54022, USA
   \newline
   (17) Division of High Energy Physics, Uppsala University, S-75121 Uppsala, Sweden
   \newline
   (18) Dept. of Physics, Stockholm University, SE-10691 Stockholm, Sweden
   \newline
   (19) Vrije Universiteit Brussel, Dienst ELEM, B-1050 Brussel, Belgium
   \newline
}
}

\maketitle

%\abstracts{}

\begin{wrapfigure}[21]{r}{0.6\columnwidth}
\begin{center}
\vspace{-10ex}
\epsfig{file=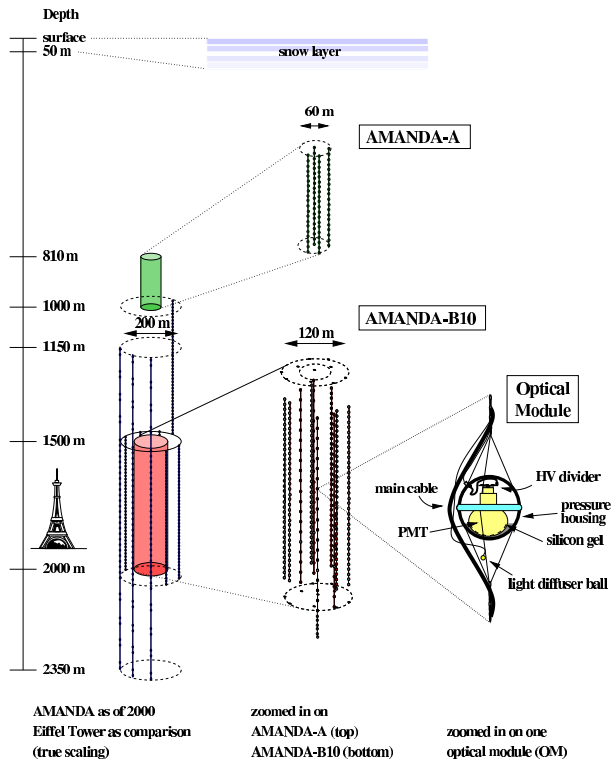,width=0.50\columnwidth}
\caption{The AMANDA detector.}
\label{Fig:AMANDA}
\end{center}
\end{wrapfigure}
\section{Introduction}
\label{Introduction}
AMANDA-II, the final stage of AMANDA before the advent of
IceCube \cite{AhrenIce}, has
been in operation since January 2000. Here, we will present results
from 131 days of the year 1997, taken with AMANDA-B10 and preliminary
results from 197 days of 2000, using AMANDA-II.

The AMANDA-II neutrino telescope, located at the geographic South
Pole, uses the ice of the antarctic ice sheet as a Cherenkov medium.
677 optical modules (OM) containing Hamamatsu R5912-2 pho\-to
multi\-pliers record the
light at depths between 1150 m and 2350 m below the surface
(fig. \ref{Fig:AMANDA}).

The prime goal of AMANDA is the detection of extraterrestrial
neutrinos. Not bound to limitations of the
two other means of observation, i.e. charged particles and gamma rays,
they arrive at the Earth uninfluenced by magnetic fields or scattering
off the Cosmic Microwave Background. The observation of a neutrino
point source and its corresponding neutrino energy spectrum would
give insight to the acceleration mechanism within the source, even if
the source is optically thick. No other particles can escape the
source without their spectra being distorted by interactions.

Source candidates can be classified into galactic and
extragalactic. Inside our galaxy, fluxes from Super-Nova Remnants
(SNR) and Micro Quasars (MQ) as well as Binary Systems have been
predicted. Most prominent Extra-Galactic sources are Active
Galactic Nuclei (AGN) and Gamma Ray Bursts (GRB) \cite{Learn,Waxma}.

%Once the neutrinos reach the Earth, a fraction of them reacts in
%weak interactions resulting in the charged lepton of the same family
%as the neutrino.
%
%In case of muon neutrinos, the emerging muon gives rise to Cherenkov
%radiation inside the glacial ice. Measuring the time and location of
%the Cherenkov cone will allow to infer the direction of the original
%neutrino.
%In case of neutral currents or electron neutrinos giving rise to
%electrons, which travel only some meters, the resulting light pattern
%is rather spherical or ellipsoidal, giving the possibility to
%distinguish them from muon neutrinos.

\section{Atmospheric Neutrinos}
The flux of atmospheric neutrinos is well known and thus provides a
means of performance checking of the detector. Muons from atmospheric
neutrinos are separated from muons produced by cosmic radiation by
determining their origin as coming from below and above.
% As a discrimination of muons
%produced by atmospheric neutrinos against the muons produced directly
%by cosmic rays is not possible, one uses the Earth as a shield
%against them.
This requires that the direction of the particles
as seen by AMANDA can be reconstructed and a sufficient background
suppression can be achieved. At the depth at which AMANDA is located, 
the flux ratio for atmospheric muons and muons from neutrinos is roughly
$10^6:1$; this gives an idea of the challenge to be conquered.

Before starting reconstruction of events, the data is scanned for
irregularities in detector behaviour. This monitoring takes place
online over the whole data-taking period. Based on the observations
made during this process, runs and OMs are excluded from the analysis.

As a first processing step, each event undergoes a
hit-cleaning which removes hits not caused by a
passing particle. These are caused by noise and after-pulsing of the
PMT, or induced in the recording electronics by induction from neighbouring
channels with a high signal level.

Then, three reconstruction levels are passed. In reconstruction, the
arrival times of the Cherenkov photons at the OMs are used to assess
the track direction and vertex. The direction is mainly
characterised by the zenith angle, where zenith angle between
$0^\circ$ and $90^\circ$
means a particle entering the detector from above (i.e. from the
South) and angles between $90^\circ$ and $180^\circ$ indicates a
particle from below (i.e. from the North).

The first reconstruction level, reducing
the experimental data to 1\%, retaining 90\% of the atmospheric
neutrino simulation, consists of a fast, non-minimising,
i.e. explicitly calculable reconstruction. It does
not take into account possible delay of the Cherenkov photons due to
scattering in the ice.
Events
reconstructed with a zenith angle bigger than $70^\circ$ are kept and
submitted to a reconstruction which corrects a track hypothesis with
a maximum likelihood method. It uses the probability of the amount of
scattering of the Cherenkov photons \cite{Wiebu}. This scattering is
due to the optical properties of the ice embedding the OMs. To evade
local peaks in the landscape formed by the mapping of the track
parameters to a likelihood, the reconstruction is iterated 16 times in
order to jump off the peaks and find the true global maximum. Events with a
resulting zenith angle of greater than $70^\circ$ are kept.

Level 2 first restricts the result of the iterative likelihood fit to
zenith angles above 80 degrees. 
More reconstructions are added fitting the spherical/ellipsoidal
light emission of the cascade rather than a track 
to determine the nature of the event: track like or cascade
like.

\begin{wrapfigure}[18]{r}{0.5\columnwidth}
\begin{center}
\vspace{-8ex}
\epsfig{file=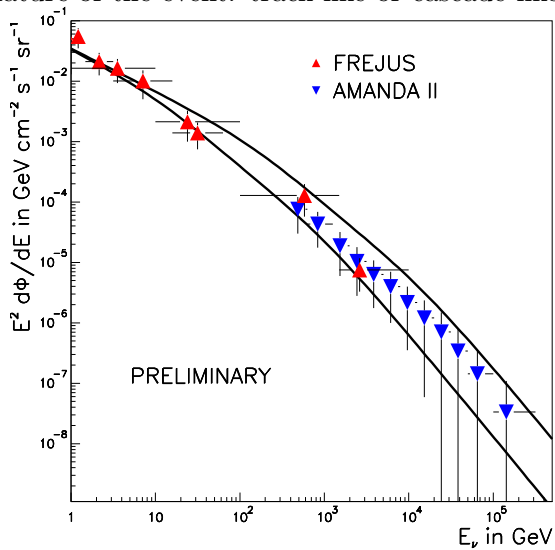,width=0.50\columnwidth}
\caption{Atmospheric neutrino spectrum as seen by AMANDA-II and
Fr\'{e}jus.}
\label{Fig:AtmNu}
\end{center}
\end{wrapfigure}
At level 3 we place restrictions on the reconstruction
results \cite{Hausc} and the topology \cite{Boese} of the event in order
to arrive at the desired background rejection.
The data is now
suppressed by a factor of $0.4\times 10^{-6}$; the signal simulation
of atmospheric neutrinos is retained to 40\%.

The resulting year 2000 data sample is thus assumed to represent mainly muons
induced by atmospheric muon neutrinos. A small background
contamination from atmospheric muons remains, especially around the
horizon, thus an additional cut on the zenith angle to be greater than
$100^\circ$ is applied.

To determine the neutrino flux, it is necessary to reconstruct the
energy of the muons. This is done by means of a neural network which
is trained on simulated events. Simple variables such as the number of
hit channels are used as input and the output is made to match the
generated energy of the particle. Then, the energy of the neutrino
itself is obtained from regularised unfolding taking into account
the energy transfer distribution from the neutrino onto the muon.

Thus one finally arrives \cite{Geene} at the spectrum shown in figure
\ref{Fig:AtmNu}. We state good agreement with the Fr\'{e}jus
observations \cite{Daum}. Non-consideration of systematic errors due to the
misreconstructed down-going muon background so far
prevents extraction of a limit on the non-atmospheric flux.

\section{Neutrino Induced Cascades}
The track based analysis of the atmospheric neutrinos is restricted to
the muonic flavour. By looking to cascades, which have a spherical
light emission pattern, one expands the capabilities to electron and
tau neutrinos. An additional benefit is that light from in-situ
sources gives rise to event patterns
similar to that of showers. This allows for verification  of the
energy resolution to
about 0.15 in $\log_{10}(E)$ in the range between 1 TeV and 100
TeV. Analysis of the year 2000 data yields an upper flux limit for the
sum of all flavours:
$\phi_{90}\times E^2 = 9\times
10^{-7}/(\mbox{GeV}^{-1}\mbox{s}\;\mbox{cm}^2\mbox{sr})$. Compared to
the AMANDA-B10 result \cite{AhrenCasc}, this is an improvement by one order
of magnitude.

\begin{wrapfigure}[18]{r}{0.5\columnwidth}
\begin{center}
\vspace{-9ex}
\epsfig{file=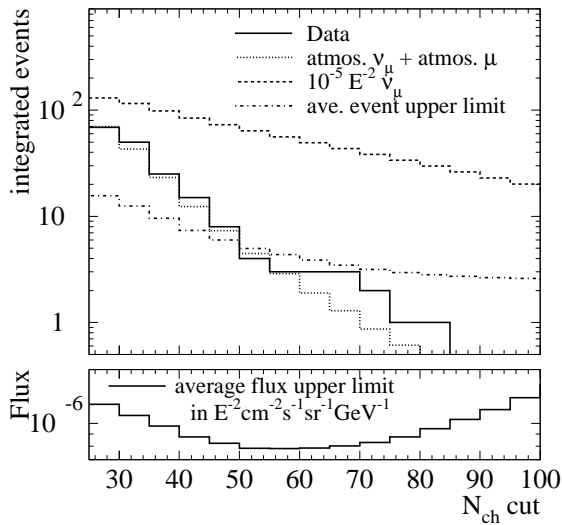,width=0.50\columnwidth}
\caption{Integrated distributions of event numbers as a function of the
number of channels cut (top). The minimum in the average flux upper
limit is found by minimising the ratio of the average upper limit to
the expected $E^{-2}$ signal.}
\label{Fig:MRF}
\end{center}
\end{wrapfigure}
\section{Diffuse Flux from Extraterrestrial Neutrinos}
We now present the search for diffuse high energy neutrino flux from
extra--terrestrial sources in 1997 data \cite{AhrenDiff}. One exploits
the harder spectrum of index -2.0 compared to the index of -3.7 for
atmospheric neutrinos. Starting from a highly enriched
neutrino sample, the number of channels is used as an
energy estimator and a cut on it applied in order to minimise the
model rejection factor (MRF \cite{Hill}), defined as the ratio of the average
upper limit and the number of expected signal events \cite{Feldm}, see
figure \ref{Fig:MRF}. An upper flux limit at 90\% C.L. of
$\phi_{90}\times E^2 = 8.4\times 10^{-7}/(\mbox{GeV}^{-1}\mbox{s}\;\mbox{cm}^2\mbox{sr})$
is obtained.

\section{Ultra High Energy (UHE) Neutrinos above 1 PeV}
At high energy, the Earth becomes opaque to UHE neutrinos because of
the rising cross section.
It is
however possible to search for high energy neutrinos above 1 PeV.
This is done by looking at events close to the horizon and
looking at the fraction OMs with only one hit, which enables to distinguish
between bundles of atmospheric muons and single muons from high energy
neutrinos. This information, together with other observables is fed
into a neural net trained to indicate the nature of the event -- muon
bundle or single muon. This yields a sensitivity \cite{Feldm} of
$E^2\phi_{\mbox{90}}=9.3\times
10^{-7}/(\mbox{GeV}^{-1}\mbox{s}\;\mbox{cm}^2\mbox{sr})$. 
The actually obtained upper flux limit for 1997 data assuming an
$E^{-2}$ spectrum (including systematics) is
$\phi_{90}\times E^2 = 7.2\times
10^{-7}/(\mbox{GeV}^{-1}\mbox{s}\;\mbox{cm}^2\mbox{sr})$. 

\section{Search for Neutrino Point Sources}
\begin{wraptable}[6]{r}{0.67\columnwidth}
\vspace{-4ex}
\begin{tabular}{lrrrrr}
Candidate & Dec. [$^\circ$] & R.A. [h] & $n_{\mbox{obs}}$ &
$n_{\mbox{bg}}$ & $\frac{\Phi_{90}}{10^{-7}/\mbox{cm}^2\mbox{s}}$ \\
\hline
\hline
Crab Nebula & 22.0 & 5.58 & 2 & 1.76 & 2.1 \\
Markarian 501 & 39.8 & 16.90 & 1 & 1.57 & 1.6 \\
Cassiopeia A & 58.8 & 23.39 & 0 & 1.01 & 1.1 \\
\end{tabular}
\caption{Neutrino flux limit $\phi_{90}$ from selected point sources}
\label{PS}
\end{wraptable}
A point source search is performed by examining a binned sky for 
an excess of events.
One calculates the significance, defined as the
negative logarithm of the probability that the observed number of
events in a bin is counted for the average of all bins.
The resulting significance distribution is compared to a distribution
obtained from random direction tracks.
For the year 2000 data, no excess in significance is observed.
Considering the background
expectation ($n_{\mbox{bg}}$) at different positions of candidate
sources (see
sect. \ref{Introduction}), we obtain the integrated fluxes above 10
GeV shown in table \ref{PS}.

\section{Neutrinos from Gamma Ray Bursts}
Gamma Ray Bursts are predicted to have neutrino emission associated
with their photon output. This allows to directly consider events
originating from the direction of the burst. The background is
measured by counting neutrino events registered 1 hour before and
after a 10
minute window around the burst time. Non observation of neutrino
excess around the 317 bursts in the 1997-2000
BATSE \cite{Batse} catalogue yields an neutrino event upper limit \cite{Feldm}
of 1.45. The calculation of a flux however requires assumptions about
the the flux normalisation \cite{WaxmaBahca} as well as taking into
account the individual nature of each GRB.

\section{Systematic Effects}
Systematics arise from uncertainties of input parameters in simulation
of particle generation and
propagation, transport of the photons through the ice and simulation
of the detector itself. For the different analyses (see table
\ref{Table:SystUncertain}), the contribution
of single uncertainties varies, but in general it can be said that the
optical properties of the ice have the biggest influence.
\begin{table}[h]
\begin{center}
\begin{tabularx}{\columnwidth}{XXXXXXXX}
Analysis & Bulk Ice & Hole Ice & OM Sens. & $\mu$ prop &
 Source\footnotemark &
Calibration \&DAQ & Total \\
\hline
\hline
Atm $\nu$, Diffuse & 15\% & 25\% & 15\% & 10\% & --- &
10\% +{$<$10\%} & 37\% \\
\hline
Cascade & 20\% & 9\% & 5\% & --- & $<$5\% & 4\% & 25\%  \\
\hline
Point Source & --- & --- & --- & --- & 25\% & --- & 25\% \\  
\hline
UHE & 34\% & --- & 12\%\footnotemark & 6\% & 8\%
(16\%\footnotemark+20\%\footnotemark) & --- &  37\%
(45\%\footnotemark)
\end{tabularx}
\caption{\label{Table:SystUncertain} Various contributions to
systematic uncertainty for different analyses. See text for
explanations.}
\end{center}
\end{table}
\addtocounter{footnote}{-5}
\stepcounter{footnote}\footnotetext{Cross-sections, Normalisations}
\stepcounter{footnote}\footnotetext{Hole ice included}
\stepcounter{footnote}\footnotetext{For
atmospheric showers due to uncertainty in
composition}
\stepcounter{footnote}\footnotetext{For atmospheric showers due to
uncertainty in the absolute flux}
\stepcounter{footnote}\footnotetext{For atmospheric showers}
%
% a separator for readability, not an empty line
%
\subsection{Primary Cosmic Ray Flux}
Absolute normalisation as well
as the element composition of the cosmic rays have uncertainties.
The first one can be estimated
by comparing the spread in the flux measured by different
experiments \cite{Hoera2002}. The composition model used \cite{Glass}
fits a heavy composition for the primaries. The effect on the UHE analysis
was obtained by inverting the proton and iron contributions as an
extreme case.

\subsection{Neutrino Cross Section}
Neutrino cross sections have been calculated \cite{Gandh} up to
$10^{21}$ eV. Below $10^{16}$ eV, all current sets of parton
distributions obtain very similar cross sections. Above, the behaviour
at Bjorken $x\rightarrow 0$ governs the result, leading to a uncertainty of up
to a factor 2 at $10^{20}$ eV, mainly important for the UHE analysis.

\subsection{Muon Propagation}
The muons produced in neutrino and cosmic ray reaction are subject to
uncertainties in the mean free path length and energy loss.
Comparison of two different muon propagation schemes,
\cite{Chirk,Lipar} allows an assessment of uncertainties here.

\subsection{Optical Ice Parameters}
The optical parameters of the bulk ice has to be determined with in-ice
devices \cite{Wosch}. Uncertainties here mainly arise
from inhomogeneities in the ice and the assumption that during
measurements photons stay within one quality of ice. We thus vary the
optical properties of ice between the observed extremes to get an
estimate on the magnitude of the effect.

\subsection{Absolute Detector Sensitivity}
The three contributions to the absolute detector sensitivity are the
sensitivity of OM itself (OM sens), which has to be combined with the
shadowing of the cables and the influence of the re-frozen water in the drill
hole (hole ice) which accommodates each OM.

\section{Conclusion}
From atmospheric neutrinos to neutrinos
of the highest energies, AMANDA shows its capabilities as neutrino
telescope. Although no other than atmospheric neutrinos are
observed, we are shown that already at limited size, it delivers
substantial results on the way to IceCube. It is however important
to further study
systematics effects, especially of the optical ice properties.
The results presented here can be found in more detail
in the cited publications as well as in the papers submitted to the
upcoming 28th International Cosmic Ray Conference, Tsukuba, Japan.

\section*{Acknowledgments}
\begin{footnotesize}
This research was supported by U.S. NSF office of Polar Programs
and Physics Division, U.S. DoE, Swedish Natural Science Research
Council, Swedish Polar Research Secretariat, Wallenberg Foundation
(Sweden), German Ministery for Education and Research, DFG (Germany),
Belgian funds for Scientific Research (FNRS-FWO), Flanders Institute to
encourage scientific and technological research in the industry (IWT),
Belgian Federal Office for Scientific,Technical and Cultural affairs
(OSTC).

\end{footnotesize}


\section*{References}
\begin{thebibliography}{99}

\begin{footnotesize}
\bibitem{AhrenIce}J. Ahrens {\it et al.}, astro-ph/0305196

\bibitem{AhrenCasc}J. Ahrens {\it et al.}, astro-ph/0206487, submitted
to {\PRD}

\bibitem{AhrenDiff}J. Ahrens {\it et al.}, astro-ph/0303218, accepted
by {\PRL}

\bibitem{Batse}http://f64.nsstc.nasa.gov/batse/grb/catalog
\bibitem{Boese}S. B\"{o}ser, Diploma Thesis, TU M\"{u}nchen, Germany, 2002

\bibitem{Chirk}D. Chirkin, and W. Rhode, Contributed to 27th
International Cosmic Ray Conference (ICRC 2001), Hamburg, Germany,
2-15 Aug 2001

\bibitem{Daum}K. Daum, \Journal{Zeitschrift f\"{u} Physik
C}{66}{417}{1995}

\bibitem{Feldm}G.J. Feldman and R.D. Cousins,
\Journal{\PRD}{57}{3873}{1998}

\bibitem{Gandh}R, Gandhi, C. Quigg, M.H. Reno and I.Sarcevic,
\Journal{\APP}{5}{81}{1996}

\bibitem{Geene}H. Geenen, Diploma Thesis, UGH Wuppertal, Germany, 2002

\bibitem{Glass}R. Glasstetter {\it et al.} [KASCADE Collaboration],
\Journal{FZKA-6345E}{}{}{}, prepared for the 26th Inernational Cosmic
Ray Conference (ICRC 1999), Salt Lake City, UT, 17-25 August 1999.

\bibitem{Hausc}T. Hauschildt, Technical Report, DESY, Zeuthen 2001

\bibitem{Hill}G.C. Hill and K. Rawlins, \Journal{Astropart. Phys}{19}{393}{2003}

\bibitem{Hoera2002}J.R. Hoerandel, astro-ph/0210453 {\it submitted to
Astroparticle Physics 15. April 2002; accepted 6. August 2002}

\bibitem{Learn}J.G. Learned and K.Mannheim, \Journal{Ann. Rev. Nucl. Part.}{50}{679}{2000}

\bibitem{Lipar}P. Lipari and T.Stanev, \Journal{\PRD}{44}{3543}{1991}

\bibitem{Waxma}E. Waxman
\Journal{Nucl. Phys. Proc. Suppl.}{100}{314}{2001}

\bibitem{WaxmaBahca}E. Waxman, J. Bahcall
\Journal{\PRL}{78}{2292}{1997}

\bibitem{Wiebu}C.H.V. Wiebusch, \Journal{DESY-Proc}{1999-01}{}{1999}

\bibitem{Wosch}K. Woschnagg, HE.4.1.15 in Proceedings of the 26th
International Cosmic Ray Conference, Salt Lake City, 1999 
\end{footnotesize}
\end{thebibliography}
\end{document}